# Wind-driven collisions between floes explain the observed dispersion of Arctic sea ice

Bryan Shaddy,[1] P. Alex Greaney,[2] and Bhargav Rallabandi[2, *]

[1]*Department of Aerospace and Mechanical Engineering,*
*University of Southern California, Los Angeles, California 90089, USA*

[2]*Department of Mechanical Engineering, University of California, Riverside, California 92521, USA*

The transport of sea ice over the polar oceans plays an important role in climate. This transport is driven predominantly by turbulent winds, leading to stochastic motion of ice floes. Observed diffusivities and velocity distributions of sea ice deviate by orders of magnitude from Brownian models, making it challenging to predict ice transport. We fully resolve these gaps through stochastic granular simulations that account for interactions between ice floes as they respond to a noisy wind. Using only directly measured quantities as model inputs, we reproduce the dispersion, diffusivity, velocity distribution, and power spectra of sea ice observed in the Fram Strait with remarkable quantitative accuracy. We understand these features as direct consequences of collisions between floes, which rapidly dissipate the energy injected by wind. A kinetic theory provides insights into these dynamics in terms of environmental properties, and makes predictions in close agreement with observations. The ideas and tools developed here pave the way for a new predictive understanding of global sea ice transport in terms of local floe-scale processes.

Sea ice occupies a significant fraction of the polar ocean surface, and its transport plays a crucial role in local ecology and global climate. The ice cover is composed of floes that are a few meters thick and between several meters and several kilometers across. Turbulent wind and ocean currents drive the motion of the floes, which consists of a mean drift and fluctuations [1, 2]. These fluctuations control spreading (or dispersion) of the ice, which plays a crucial role in the global transport of sea ice. Observations have revealed unusual features of ice fluctuations, such as fat-tailed velocity distributions and anomalous dispersion [3–5]. The lack of understanding of these features poses challenges to the prediction of ice transport in the polar oceans. Acute changes in ice properties and environmental conditions in recent years amplify these challenges, leading to uncertainties in climate forecasts. Here, we show how the interplay between environmental noise and interactions between floes resolves observations of sea ice transport with quantitative accuracy.

## UNEXPLAINED ICE TRANSPORT

Wind is the primary driver of ice motion in the Arctic, while ocean currents usually play a smaller role. Fluctuations in Arctic wind speed are about 5 m/s with correlations over timescales of about 30 hours (Supplementary Information [6]), driving ice floes to fluctuate at speeds of 10 cm/s as they drift. Isolated floes adjust rapidly (on timescales of about 1 hr) to environmental forcing and are thus slaved to the noisy wind. With these scales, and assuming the ice behaves like a Brownian particle [7] we estimate a diffusivity of $(10\,\mathrm{cm/s})^2 \times 30\,\mathrm{hr} \approx 1000\,\mathrm{m^2/s}$. However, observed diffusivities are between 20–100 m²/s [5], a full order of magnitude smaller. Furthermore, the distribution of ice velocity is narrower than expected from the normally distributed wind fluctuations. Meanwhile, ice velocity distributions are non-Gaussian and have overpopulated tails [1, 4, 5, 8–10]. Observed power spectra are also flatter than expected from a Brownian process, containing proportionately more energy at higher frequencies. These discrepancies between observations and models have been attributed to the intermittency of the wind [11], the turbulent structure of wind or ocean currents [5, 12], long-range spatial correlations [1], and intermittent ice fractures [13].

We show here that all these puzzles are *simultaneously* and quantitatively resolved with one key physical ingredient: interactions between floes as they respond to noisy winds. Ice floes are rarely isolated, and can cover as much as 80–100% of the ocean surface in polar regions, making floe-floe interactions inevitable. These interactions are essential in the rheology of the ice pack [14], but their effect on fluctuations and dispersion of the ice has largely been ignored. We will see that interactions occur frequently, and on timescales much shorter than those characterizing wind noise. These interactions dissipate energy to produce narrower velocity distributions, and lead to shorter mean-free paths, giving rise to smaller diffusivities and flatter power spectra.

## STOCHASTIC AIR-ICE-OCEAN MODEL

We develop a stochastic discrete element model (SDEM) that resolves the dynamics of interacting floes which are transported over a two-dimensional ocean by noisy winds (Fig. 1a); see also [15–17]. Mimicking the conditions in the Arctic, we consider floes with a wide distribution of sizes, occupying an area fraction $\varphi$; see Fig. 1b. The velocity $\mathbf{v}_i$ of a floe $i$ is driven by stresses due to a wind velocity $\mathbf{u}_i$ on its top surface, and resisted by an ocean drag on its bottom surface. We model floe interactions as energy-dissipating collisions between rigid

* bhargav@engr.ucr.edu

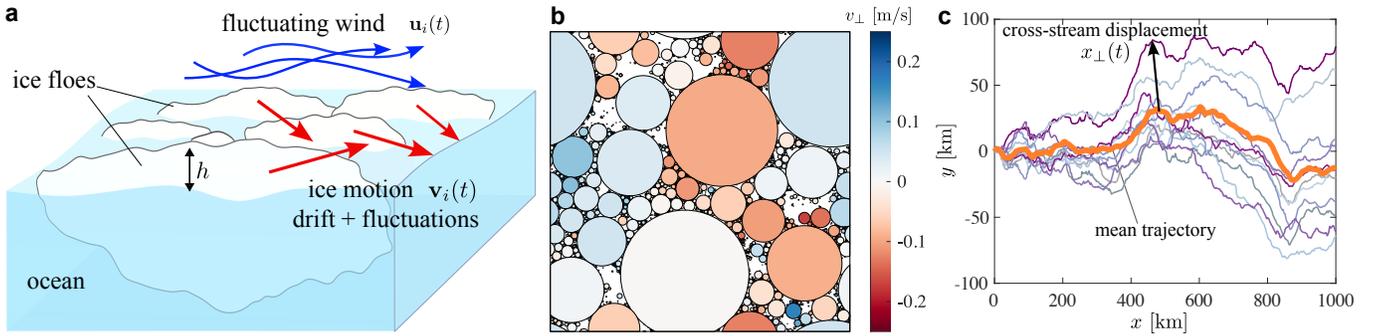

FIG. 1. (a) Fluctuating winds drive the motion of interacting ice floes on the ocean surface. (b) Snapshot of simulation showing floes, colored by their cross-stream fluctuating velocities $v_\perp$. (c) Sample floe trajectories (blue-purple), indicating a mean trajectory (orange), and cross-stream fluctuations.

floes that occur on contact. Between collisions, conservation of momentum of floe $i$ reads [1]

$$\rho h \frac{d\mathbf{v}_i}{dt} = \rho_a C_a |\mathbf{u}_i| \mathbf{u}_i - \rho_w C_w |\mathbf{v}_i| \mathbf{v}_i, \qquad (1)$$

and depends on the density $\rho$ and thickness $h$ (about a meter) of the floe, the densities $\rho_a$ and $\rho_w$ of the air (wind) and water (ocean), and respective drag coefficients $C_a$ and $C_w$. These quantities are well characterized for typical conditions and are tabulated in the Supplementary Information (SI; [6]). Ocean currents, Coriolis forces and tidal effects typically have a weaker effect on ice transport in the Arctic [1, 18]. Without fluctuations, the ratio of the steady-state (free-drift) ice speed to the wind speed is given by the Nansen number $\mathcal{N} = (\rho_a C_a / \rho_w C_w)^{1/2} \approx 0.02$. Floes respond to changes in wind velocity over an inertal timescale $\tau_d = \rho h / (\bar{u} \sqrt{\rho_a \rho_w C_a C_w}) \approx 1$ hr, where $\bar{u}$ is the mean wind speed.

We find that only a handful of properties of the turbulent atmosphere are necessary to understand ice motion. To good approximation, wind fluctuations are isotropic, normally distributed with a standard deviation $\sigma$, and exponentially decorrelated with time constant $\tau$ [1, 6]. Additionally, a mean wind velocity $\bar{\mathbf{u}}$ drives the steady drift of the ice. The simplest description of a stochastic wind with these properties has a velocity governed by the Langevin equation [19–21]

$$\frac{d\mathbf{u}_i}{dt} = \frac{1}{\tau}(\bar{\mathbf{u}} - \mathbf{u}_i) + \sqrt{\frac{2\sigma^2}{\tau}} \boldsymbol{\eta}_i(t), \qquad (2)$$

where $\boldsymbol{\eta}_i(t)$ is a two-dimensional white noise. Equation (2) is thus a parsimonious representation of the complex turbulent atmosphere.

The final ingredient accounts for interactions between floes, which involve elastic and plastic deformation of the ice to form pressure ridges, finger rafts, or fractures, and may ultimately lead to their fragmentation or coalescence [1, 18, 22, 23]. We model these processes as energy-dissipating (inelastic) but momentum-conserving collisions between floes. We treat each floe $i$ as a rigid disk of radius $a_i$, interacting through collisions characterized by a coefficient of restitution $r$ between 0 and 1 [15]. The velocity of floe $i$, upon colliding with floe $j$, becomes [24]

$$\mathbf{v}_i \xrightarrow{j} \mathbf{v}_i - \frac{a_j^2}{a_i^2 + a_j^2}(1+r)(\mathbf{v}_i - \mathbf{v}_j) \cdot \mathbf{n} \mathbf{n}, \qquad (3)$$

where $\mathbf{n}$ is the unit normal between the floes at the instant of contact. Floe sizes typically exhibit a self-similar (power-law) distribution. Consistent with observations [25–27] and recent computational studies [28], we draw floe radii from a size distribution $P_s(a) \propto a^{-2}$, between 10 m and 4500 m; see Fig. 1b. While the cutoffs of the distribution set the length scales of the system, the polydispersivity is itself not crucial as we later discuss. We neglect melting or freezing as they only become appreciable on timescales (months) much longer than those of interest here (days).

## MODEL PREDICTIONS AND COMPARISON WITH OBSERVATIONS IN THE FRAM STRAIT

We simulate the motion of $N = 2000$ (large enough to ensure statistical convergence) ice floes at a fixed area fraction $\varphi$, by solving (1)–(2) in a periodic domain of side length $L = (\sum_i \pi a_i^2 / \varphi)^{1/2}$; see *Methods* A. We inform the model inputs with statistics gathered from the ERA5 wind dataset for the Fram Strait during the winter months of January, February, and March for the years 2002–2003 and 2007–2009. These wind data follow the geographical region and time periods over which the FRAMZY/ACSYS sea ice dataset was analyzed by [5]. From the ERA5 wind data [29], we find a mean wind speed $\bar{u} \approx 4.8$ m/s, with isotropic fluctuations with a standard deviation of $\sigma \approx 6.0$ m/s correlated over a time of $\tau \approx 32$ hours. The data analysis to extract these quantities is detailed in the SI [6]. All model parameters are obtained directly from existing field, satellite or reanalysis data, with the sole exception of the coefficient of restitution $r$, which we set to 0.1 throughout. The effects of varying $r$ are modest and are discussed in the SI [6].

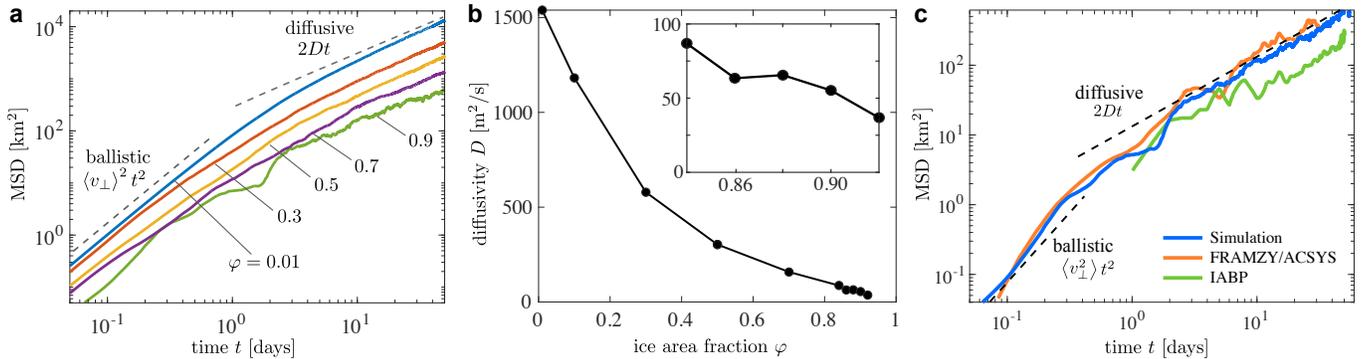

FIG. 2. (a) Simulated mean-squared displacement (MSD) in the cross-stream direction as a function of time for different ice area fractions $\varphi$. After an initial ballistic regime, the dispersion is diffusive, i.e. $\langle x_\perp^2 \rangle \sim 2Dt$. (b) Diffusivity $D$ decreases sharply with ice area fraction $\varphi$; the inset shows data for $\varphi$ between 0.84 and 0.92. (c) Simulated MSD at $\varphi = 0.9$ is in excellent agreement with observations (data taken from [5]) over three decades in time.

Simulations yield positions $\mathbf{x}_i(t)$ and velocities $\mathbf{v}_i(t)$ of each floe $i$ (Fig. 1b,c). Floes drift with a mean velocity (due to the mean wind), and fluctuate both along and across the drift direction. We focus on *cross-stream* displacements $x_\perp(t)$ and velocities $v_\perp(t)$ of the floes, which are well-characterized in observations. We extract cross-stream fluctuations from the simulated trajectories and then compute various macroscopic transport properties (see *Methods* B), which we compare directly with the ice observations analyzed by [5] in the Fram Strait for a region and time consistent with the model inputs. The observed ice data exhibit some variation of the ice fraction, with an average of about 0.9, so we consider a range of ice area fractions. We also include a comparison with the IABP ice data in the central Arctic [3–5] which displays similar trends to those in the Fram Strait.

### Cross-stream dispersion

The cross-stream mean-squared-displacement (MSD) $\langle x_\perp \rangle^2$ (angular brackets denote an ensemble average) describes the spread, or dispersion, of ice floes away from the mean trajectory (Fig. 1c). The dispersion is ballistic (MSD $\sim \langle v_\perp^2 \rangle t^2$) at short times. It transitions to diffusive behavior at long times (MSD $\sim 2Dt$), and is characterized by a cross-stream diffusivity $D$ (Fig. 2a). The MSD decreases sharply as $\varphi$ increases (note the logarithmic scales in Fig. 2a). Fig. 2b highlights this by plotting diffusivity against ice area fraction. At small ice fractions we find $D \approx 1580 \, \text{m}^2/\text{s}$, which agrees with the theoretical prediction for isolated floes, $D = (\mathcal{N}\sigma)^2 \tau$, but is an order of magnitude greater than observed diffusivities for the Fram Strait. Increasing the area fraction leads to a sharp decrease in $D$, as collisions reduce the mean-free path of the floes.

At $\varphi = 0.9$, representative of the FRAMZY/ACSYS observations [5], the model achieves a remarkable *quantitative* reproduction of the observations for the MSD and, as we will later show, other measures of stochastic ice transport. Figure 2c plots the cross-stream MSD, showing excellent agreement between simulations and observations over 3 decades of timescales, ranging from one hour to one month. From the MSD, we extract diffusivities in the range 40–100 m$^2$/s over a range of ice concentrations $0.84 < \varphi < 0.92$ (similar values are obtained from the power spectrum [6]). These predictions are again consistent with observed diffusivities, and their range mirrors the variability in the observations, e.g., due to daily variations in ice concentration [5].

### Velocity distribution

The stochastic wind and collisions between floes lead to a distribution of cross-stream velocities, which varies with area fraction $\varphi$ (Fig. 3a). We recall that the ice dynamical time $\tau_d \approx 1$ hr is an order of magnitude smaller than the wind correlation time $\tau$. At small ice fractions, collisions are infrequent and so floes adjust rapidly to fluctuations of the wind. In this regime we expect Brownian-like dynamics (*Methods* C) leading to $\langle v_\perp^2 \rangle^{1/2} \approx \mathcal{N}\sigma \approx 12$ cm/s, which is consistent with the simulations. The velocity PDF narrows as $\varphi$ increases, as made clear by Fig. 3b which plots the root-mean-squared (RMS) velocity $\langle v_\perp^2 \rangle^{1/2}$, which characterizes the width of the PDF, against $\varphi$. Collisions are more frequent at large $\varphi$, dissipating energy and thereby leading to slower motion. The inset of Fig. 3b shows that collisions dissipate a greater fraction of the work done by wind at large $\varphi$ (the rest is dissipated by ocean drag, see *Methods* B).

Additionally, the PDF significantly changes in *shape* with $\varphi$ (Fig. 3c; note the logarithmic vertical axis). The distribution is Gaussian ($\propto e^{-kv_\perp^2}$) at small $\varphi$, reflecting the distribution of wind velocities, and consistent with a Brownian process. As $\varphi$ increases, the PDF is no longer Gaussian but instead exponential ($\propto e^{-q|v_\perp|}$), mirroring

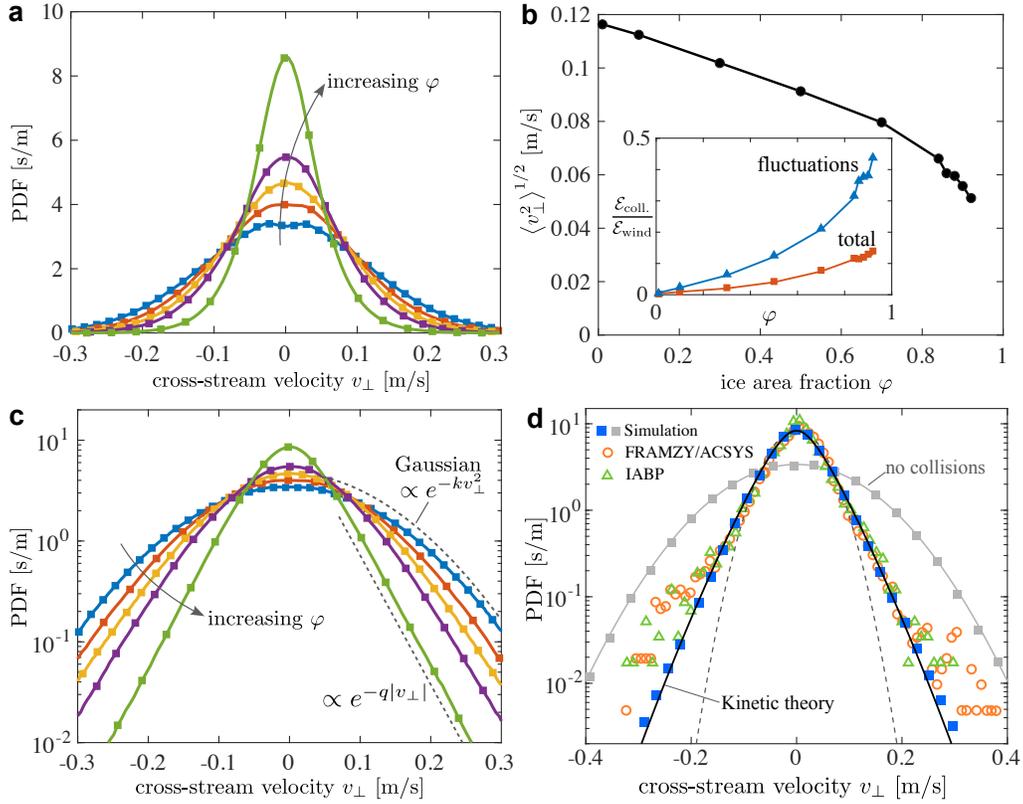

FIG. 3. (a) Cross-stream ice velocity PDF narrows with increasing ice fraction $\varphi$. (b) The width of the distribution decreases with $\varphi$. Inset shows the fraction of the energy injected by the wind that is dissipated by collisions, which increases with $\varphi$. (c) Same data as (a) plotted on logarithmic axes, showing a transition between a Gaussian and an exponential distribution at larger area fractions. (d) Simulations in the Fram Strait at $\varphi = 0.9$ (blue symbols) are in excellent agreement with observations in the Fram Strait [5] (orange symbols) and predictions of a kinetic theory (solid curve). These results are simultaneously much wider than the best-fit Gaussian (dashed curve), and much narrower than the Gaussian PDF without collisions (gray symbols and curve).

observations of sea ice [3, 5, 13] and studies of wet granular systems [30–33].

With the same parameters as in Fig. 2c, the simulations capture the observed velocity PDF in the Fram Strait [5], with excellent quantitative agreement across 3 orders of magnitude (Fig. 3d). This distribution is much narrower than in the dilute limit $\varphi \to 0$ (gray symbols and solid curves), while simultaneously being much wider than the best-fit Gaussian (dashed curve). The model yields an RMS velocity $\langle v_\perp^2 \rangle^{1/2} \approx 5.96$ cm/s, within about 5% of the observations.

**Power spectral density**

Finally, we quantify the temporal characteristics of the ice motion by studying the power spectral density (PSD), which quantifies the cross-stream fluctuation kinetic energy per frequency $f$ (see *Methods* B). Figure 4 plots the PSD, once more showing excellent agreement between the simulations (with the same model inputs as before) and observations across 2 decades of frequency, corresponding to time scales ranging between a few hours to about 10 days. The sharp peak in the observations at $(12\,\mathrm{hr})^{-1} \approx 2.3 \times 10^{-5}$ s$^{-1}$ is likely from tidal effects, which we do not model. The PSD scales roughly as $f^{-1}$ over a wide range of frequencies, which transitions to an $f^{-2}$ scaling at higher frequencies, consistent with previous observations [5, 13]. The transition occurs at $\approx 7 \times 10^{-5}$ Hz (time scales of a few hours), close to the ice dynamical timescale $\tau_d$. The PSD in the absence of interactions is noticeably different from the observations, greatly overpredicting it at low frequencies and underpredicting it at higher frequencies. Collisions flatten the spectrum by decreasing the diffusivity (low frequency behavior) while boosting high-frequency contributions due to the shorter mean-free paths of the floes.

**DISCUSSION**

The simulations do a remarkable job of reproducing a number of quantitative measures of ice transport. We reiterate that all model inputs in the simulations are ob-tained directly from existing observational data. We find that it is essential to account for floe-floe interactions to recover observed statistics of ice motion. It is also important that the driving wind fluctuations occur on much longer timescales than collision timescales, which allows floes to experience many collisions before the wind changes direction.

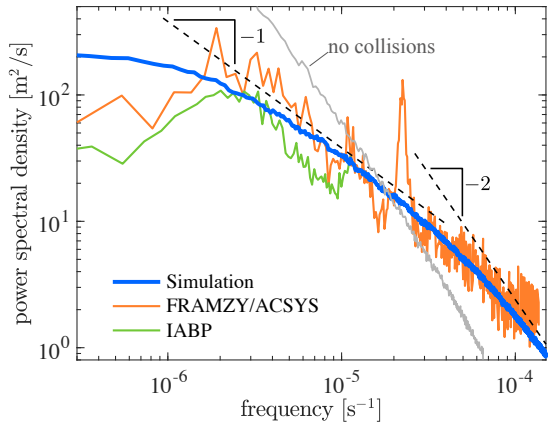

FIG. 4. Power spectral density for sea ice motion for the Fram Strait, showing excellent agreement between simulations and observations [5] over 2 decades of frequency. Slopes of $f^{-1}$ and $f^{-2}$ are indicated. Neglecting collisions overpredicts the PSD at low frequencies and underpredicts it at higher frequencies.

Other features play smaller roles in the ice transport. For example, polydispersity of floe sizes is not necessary for the model to reproduce the observations. We observe little quantitative difference in the model predictions when the power-law floe size distribution is replaced by a single equivalent floe radius of 750 m (SI [6]), which approximately preserves the surface-area-to-volume ratio of the full size distribution (*Methods* D). We also find that the coefficient of restitution $r$ has modest effects on transport properties (SI [6]). Ocean eddies have been shown to underpredict ice fluctuations [12], though they are important to ice rotation [34].

### Kinetic theory

To better understand the predictions of the SDEM, we develop a coarse-grained kinetic theory, following the arguments of [32, 35, 36] for wet granular media but accounting for stochastic forcing due to wind. We start from a Boltzmann transport equation (BTE) for the cross-stream velocity distribution $P(v_\perp, t)$ in a population of floes making up a spatially uniform ice field. The BTE describes the time-rate of change of the velocity distribution due to (i) wind noise, which injects energy, (ii) ocean drag, which decelerates floes, and (iii) floe-floe collisions, which redistributes velocities while conserving momentum but dissipating energy. The rate of change of probability due to collisions $(\partial P/\partial t)_{\text{coll.}}$ is a combination of a gain rate $G$ and a loss rate $L$, which we model as linear in the distribution (detailed derivation in *Methods* C). Floes are removed from the distribution at a rate $L = \nu P(v_\perp, t)$, where $\nu$ is a collision rate with units of inverse time. Collisions dissipate energy, causing floes to slow down on average. We use $\alpha^{-1}$ to denote the factor by which a floe slows down upon one collision; energy conservation arguments find that $\alpha = 2/(1 + r^2) = 1.98$. Then, a floe traveling with velocity $\alpha v_\perp$ will produce floes with velocity $v_\perp$ post-collision ($\alpha$ such floes are produced to conserve momentum). This yields a gain rate $G = \nu \alpha P(\alpha v_\perp)$. The rate of change of probability due to collisions is therefore $(\partial P/\partial t)_{\text{coll.}} = G - L = \nu(\alpha P(\alpha v_\perp) - P(v_\perp))$. With this collision model, the BTE at steady state reduces to a differential equation governing $P(v_\perp)$,

$$(\mathcal{N}\sigma)^2 \frac{\mathrm{d}^2 P(v_\perp)}{\mathrm{d} v_\perp^2} + \frac{\mathrm{d}(v_\perp P(v_\perp))}{\mathrm{d} v_\perp} + \nu\tau[\alpha P(\alpha v_\perp) - P(v_\perp)] = 0. \quad (4)$$

The first and second terms account for the wind noise and ocean drag, while the third term is the effect of collisions.

The collision rate $\nu$ is the inverse of a typical collision time, and depends on the mean-free-path of a floe and the speed of fluctuations. Low concentrations correspond to the limit $\nu = 0$, and (4) recovers a Gaussian PDF (thin gray curve in Fig. 3d) with a standard deviation ($\mathcal{N}\sigma$), fully capturing simulations (gray symbols in Fig. 3d). At greater area fractions, the collision term becomes important and the Boltzmann transport theory yields narrower distributions with exponential tails. From the SDEM model inputs, we estimate a collision rate $\nu \approx 0.31\,\text{hr}^{-1}$ using geometric arguments (*Methods* C and D). If we instead treat the collision rate as a free parameter, we find that solutions of (4) best fit the simulated PDF at $\nu = 0.27\,\text{hr}^{-1}$, which is rather close to this estimate. With this best-fit $\nu$, the theory (4) (solid curve in Fig. 3d) is in remarkable agreement with the observed and simulated ice velocity PDFs over three decades.

For $\varphi = 0.9$, the time between collisions, $\nu^{-1} \approx 3.4\,\text{hr}$ is much shorter than the wind decorrelation time $\tau \approx 32\,\text{hr}$. Analyzing (13) in the regime $\nu\tau \gg 1$ reveals that collisions (as opposed to wind drag) are the primary source of energy dissipation (*Methods* C). The PDF becomes exponential and is given by $P(v_\perp) = \frac{q}{2}\exp\{-q|v_\perp|\}$, with $q = \frac{\sqrt{\nu\tau-1}}{\mathcal{N}\sigma}$. From this distribution we obtain an RMS ice velocity

$$\langle v_\perp^2 \rangle^{1/2} = \sqrt{\frac{2}{\nu\tau - 1}}\mathcal{N}\sigma = \sqrt{\frac{2\rho_a C_a}{\rho_w C_w(\nu\tau - 1)}}\sigma, \quad (5)$$

solely in terms of environmental properties and the ($\varphi$-dependent) collision frequency. For the system under consideration, (5) yields $\langle v_\perp^2 \rangle^{1/2} \approx 5.99\,\text{cm/s}$, in excellent agreement with observations, SDEM simulations and numerical solutions of (4); cf. Fig. 3b at $\varphi = 0.9$. To close the loop, we reinterpret the motion of a floe in a concentrated ice field as Brownian motion with the new

RMS velocity $\sqrt{2/(\nu\tau - 1)}\mathcal{N}\sigma$ and relaxation time $\nu^{-1}$ (as opposed to the RMS velocity $\mathcal{N}\sigma$ and relaxation time $\tau$ of the dilute limit). This yields a prediction for the diffusivity (*Methods* C)

$$D = \frac{\langle v_\perp^2 \rangle}{\nu} = \frac{2\mathcal{N}^2\sigma^2}{\nu(\nu\tau - 1)} = \frac{2\rho_a C_a \sigma^2}{\rho_w C_w \nu(\nu\tau - 1)}. \quad (6)$$

With the relevant model parameters, (6) predicts $D \approx 52\,\text{m}^2/\text{s}$, once more reproducing the result of simulations ($55\,\text{m}^2/\text{s}$) at $\varphi = 0.9$.

## CONCLUSIONS

We have thus shown that the observed features of sea ice motion emerge as a direct and robust consequence of a noisy wind that drives interactions between floes. In concentrated ice fields, collisions are much more frequent than fluctuations of the wind. These rapid collisions drain energy from the floes, narrowing velocity distributions, lowering dispersion coefficients, and flattening power spectra relative to dilute (non-interacting) ice fields. A mean-field kinetic theory yields insight into the velocity distribution and the diffusivity in terms of a mean collision rate.

The modeling paradigm recovers observations quantitatively using direct empirical data as model inputs. Furthermore, we find that it is not necessary to appeal to large scale spatial structure of the ice field, nor any detailed notion of atmospheric or oceanic turbulence. Our findings underscore that ice motion is controlled by "local" driving mechanisms, and can thus be understood in terms of a small number of local environmental properties and processes.

These ideas constitute important novel advances in the understanding of sea ice transport processes. Further developments in floe-scale dynamics are likely to be important in predicting other transport properties of sea ice on global scales. One example is the rheology of sea ice, which is a poorly understood yet crucial part of climate models. It is conceivable that the effective ice rheology — a direct consequence of floe-floe interactions — may not be a property of the ice pack alone, but rather a consequence of how the ice interacts with its noisy environment. Making these connections across scales would lead to new insights into sea ice transport, as well as an improved understanding and forecasting of climate.

## ACKNOWLEDGMENTS

B. R. thanks Monica M. Wilhelmus for preliminary discussions. The authors thank the Hellman Family Foundation for support. B. R. acknowledges the UCR Regents Faculty Fellowship for partial support. B. S. acknowledges support from Chancellor's Research Fellowship at UCR, and the Viterbi School of Engineering/Graduate School Fellowship for Incoming Students.

## METHODS

### A. SDEM simulations

We initialize the ice field by placing floes at a random selection of points in a square domain of side $L$. The radii of the floes at these points are randomly sampled from a power law distribution, which invariably leads to (unphysical) overlaps between floes. If a pair of floes are closer than the sum of their radii, we displace each by a small amount away from each other along their line-of-centers. This process is repeated until all overlaps are resolved, leading to a physically consistent non-overlapping initial ice field. With no loss of generality, we orient the $x$ axis along the mean wind $\bar{\mathbf{u}}$, which is held constant throughout. We initialize the wind field over each ice flow, $\mathbf{u}_i$, as a superposition of $\bar{\mathbf{u}}$ and an isotropic pseudo-random noise with standard deviation $\sigma$. The velocity of each floe is then initialized to zero.

Given the wind acting on floe $i$ at time level $(n)$, $\mathbf{u}_i^{(n)}$, we integrate (2) through a time step $\Delta t$ according to

$$\begin{aligned}\mathbf{u}_i^{(n+1)} = &\bar{\mathbf{u}} - (\bar{\mathbf{u}} - \mathbf{u}_i^{(n)})e^{-\frac{\Delta t}{\tau}} \\ &+ \sigma(1 - e^{-\frac{2\Delta t}{\tau}})^{\frac{1}{2}}\mathbf{N}[0,1].\end{aligned} \quad (7)$$

Here, $\mathbf{N}[0,1]$ is a standard normal random variable (zero mean, unit variance) in two-dimensions. Given the wind velocities on each floe at time levels $(n)$ and $(n+1)$, we integrate the ice dynamics (1) through the time step $\Delta t$ (temporarily assuming no collisions) using a "semi-implicit" Euler integration to obtain floe velocities at time level $(n+1)$, $v_i^{(n+1)}$. We then integrate the velocities at time levels $(n)$ and $(n+1)$ to obtain floe positions at time level $(n+1)$. After each time step, we evaluate the distance from each floe to all other floes. If the distance between two floes is less than the sum of their radii, the collision model is implemented. Post-collision positions and velocities are then computed based on the size (mass), speed, and direction of travel of the colliding floes, as described in (3). Updated post-collision positions (non-overlapping) and velocities are then obtained, thus accounting for both winds and collision effects over a time-step. A coefficient of restitution of 0.1 was used for all the results in the main text [37]. Other values are studied in Supplementary Information [6].

Since we initialize the system with stationary floes, the initial conditions do not represent a statistical equilibrium. Thus, we first perform a "spin-up" to arrive at a statistical equilibrium. We then reset time to zero and use velocities and positions from the statistically equilibrated state to begin a new simulation, the results of which we analyze.

Our floe simulations are performed in a periodic box of side length $L$, so positions are tracked modulo this length. To accurately compute the MSD (which may become greater than $L$), we also track the number of turns each floe makes around the box along each coordinate axis. This lets us reconstruct the absolute displacement of each floe, which we use for dispersion statistics.

### B. Calculation of transport properties

The cross-stream mean squared displacement is the ensemble average of the cross-stream displacement:

$$\text{MSD}(t) = \langle x_\perp^2 \rangle = \frac{1}{N} \sum_{i=1}^{N} \left(x_{\perp,i}(t)\right)^2. \tag{8}$$

The cross-stream velocity probability density function (PDF) was obtained from the time series data $v_{\perp,i}(t)$, and is normalized to have unit integral.

From a time series $v_{\perp,i}(t)$, we first shift the time vector of the data by $T/2$ ($T$ is the maximum simulated time), so that it is centered at 0. Its Fourier transform over time $T$ is defined by

$$\hat{v}_{\perp,i}(f;T) = \int_{-T/2}^{T/2} v_{\perp,i}(t) e^{-i2\pi ft} dt, \tag{9}$$

using fast Fourier transforms (FFT). Using the transformed velocity, we compute the power spectral density (PSD) according to

$$\text{PSD}(f) = \lim_{T \to \infty} \frac{1}{T} \left\langle |\hat{v}_\perp(f;T)|^2 \right\rangle. \tag{10}$$

Note that $\int_{-\infty}^{\infty} \text{PSD}(f) df = \langle v_\perp^2 \rangle$ by definition. We utilize this fact to rescale the PSD of [5] for comparison with the present results (a different normalization was used in that article).

At statistical steady state, the kinetic energy of the collection of floes remains constant with time. Thus, the rate of work done on the floes by the wind ($\mathcal{E}_\text{wind}$) must be equal to the rate of energy dissipated by the ocean drag ($\mathcal{E}_\text{ocean}$) plus the rate of energy lost to collisions ($\mathcal{E}_\text{coll.}$). Taking a dot product of wind and ocean forces on a floe $i$ with its momentum $\rho \pi a_i^2 \mathbf{v}_i$, and adding contributions from all floes gives

$$\mathcal{E}_\text{wind} = \sum_i \rho_a C_a |\mathbf{u}_i| \mathbf{u}_i \cdot (\rho \pi a_i^2 \mathbf{v}_i) \quad \text{(input)} \tag{11a}$$

$$\mathcal{E}_\text{ocean} = \sum_i \rho_a C_a |\mathbf{v}_i| \mathbf{v}_i \cdot (\rho \pi a_i^2 \mathbf{v}_i) \quad \text{(dissipation)} \tag{11b}$$

$$\mathcal{E}_\text{coll.} = \mathcal{E}_\text{wind} - \mathcal{E}_\text{ocean} \quad \text{(dissipation)}. \tag{11c}$$

The (rates of) energy associated with the average motion of the floes (uniform drift) and the mean wind are obtained by replacing the wind and ice velocities by their ensemble averages. Subtracting the energy due to mean flow from the total, we obtain the energy associated with fluctuations.

### C. Boltzmann transport theory

To develop a kinetic theory we consider a population of floes of uniform size that are distributed uniformly in space, but with a nontrivial distribution of velocities. We first analyze transport in the absence of collisions subject to a stochastic wind, governed by (1) and (2). To simplify this set of equations, we linearize (1) by replacing $|\mathbf{u}|$ and $|\mathbf{v}|$ with their mean values, $|\bar{\mathbf{u}}|$ and $\mathcal{N}|\bar{\mathbf{u}}|$, respectively. Then, (1) yields simplified ice dynamics (suppressing subscripts identifying individual floes) $d\mathbf{v}/dt = (\mathcal{N}\mathbf{u} - \mathbf{v})/\tau_d$. As noted in the main text $\tau_d \ll \tau$, so the ice is in quasi-equilibrium with the wind on long timescales, i.e. $\mathbf{v} \approx \mathcal{N}\mathbf{u}$. We substitute this relation into (2) and subtract off the mean part to obtain a stochastic differential equation (SDE) for ice velocity fluctuations $\mathbf{v}'$

$$\frac{d\mathbf{v}'}{dt} \approx -\frac{\mathbf{v}'}{\tau} + \sqrt{\frac{2\mathcal{N}^2\sigma^2}{\tau}} \boldsymbol{\eta}(t), \tag{12}$$

This is a Langevin equation [19] that describes an isolated floe as a Brownian particle with RMS velocity $\mathcal{N}\sigma$ and diffusivity $\mathcal{N}^2\sigma^2\tau$ along any direction.

As the fluctuations are isotropic, we focus on the cross-stream fluctuations $v_\perp$, which satisfies the one-dimensional (1D) version of (12) between collisions. Accounting for collisions introduces an additional force in (12) that results from interactions with other floes. However, an expression for such a force is not straightforward to formulate. We instead develop a coarse-grained kinetic theory that governs the probability density $P(v_\perp, t)$ describing the distribution of cross-stream velocities $v_\perp$ at time $t$. Applying standard techniques from stochastic calculus [19, pp. 147] to the 1D version of (12) yields a partial differential equation governing $P(v_\perp, t)$,

$$\frac{\partial P}{\partial t} = \frac{\partial}{\partial v_\perp}\left(\frac{v_\perp P}{\tau}\right) + \frac{\mathcal{N}^2\sigma^2}{\tau}\frac{\partial^2 P}{\partial v_\perp^2} + Q[v_\perp]. \tag{13}$$

The first term on the right-hand side is due to ocean drag, and the second is due to wind fluctuations. The last term on the right-hand side of (13) is the *collision operator* $Q[v_\perp] = (\partial P/\partial t)_\text{coll.}$ which accounts for the redistribution of floe velocities due to collisions. Equation (13) is a Boltzmann transport equation (BTE), generalized to account for a stochastic wind forcing. The absence of collisions ($Q = 0$) recovers the Fokker–Planck equation, while the absence of wind fluctuations ($\sigma = 0$) recovers the classical BTE [24].

The collision operator consists of two rates: (i) a rate of loss $L$ of floes with velocity $v_\perp$ because they collide with other floes and are assigned a new velocity, and (ii) a rate of gain $G$ from floes acquiring velocity $v_\perp$ as a result of collision. The loss term is an integral over the velocity distribution weighted by the collision rate, $R(v_\perp - v'_\perp)$, for floes with approach velocity $v_\perp - v'_\perp$, and the gain

term is a double integral so that [38]

$$\begin{aligned} Q &= -L + G, \\ L &= P(v_\perp) \int R(v_\perp - v'_\perp) P(v'_\perp) dv', \\ G &= \int\int \Gamma_{v'_\perp, v''_\perp \to v_\perp} R(v'_\perp - v''_\perp) P(v'_\perp) P(v''_\perp) dv' dv'', \end{aligned}$$

where $\Gamma_{v'_\perp, v''_\perp \to v_\perp}$ is the probability that floes colliding with velocity $v'_\perp$ and $v''_\perp$ result in one of those floes having velocity $v_\perp$ post collision.

Using the full collision operator makes the BTE a nonlinear integro-differential equation that is difficult to solve, even in one dimension, so instead we replace it with a mean field approximation. We adapt the argument of [35, 36] and consider the most likely interaction: that of a fast floe of velocity $v_\perp$ with a much slower (effectively stationary) floe. In this setting the integral in the loss term $L$ can be replaced with an effective total collision rate, $\nu$, so that $L = \nu P(v_\perp)$. To simplify the gain term, we recognize that collisions drain energy from the particles while conserving momentum. We denote the factor by which a collision slows down the faster floe by $\alpha^{-1}$, where $\alpha > 1$. Then, a floe with a pre-collision velocity $\alpha v_\perp$, upon collision, leads to the gain of particles of velocity $v_\perp$. The number of such particles gained per collision is $\alpha$, since momentum is conserved. With a collision rate $\nu$ as before, we obtain the gain rate $G = \nu \alpha P(\alpha v_\perp)$. Putting these approximations together gives the simplified form of the collision operator:

$$Q[v_\perp] = -\nu P(v_\perp) + \nu \alpha P(\alpha v_\perp). \quad (14)$$

The integral of $Q[v_\perp]$ over $v_\perp$ is zero and so the collision operator is conservative. At steady state, (13) with (14) reduces to (4).

In the setting described above (a fast floe colliding with a stationary one), the post-collision energy is a factor $(1+r^2)/2$ of the pre-collision energy, $r$ being the coefficient of restitution. Meanwhile, a collision in the simplified model generates an $\alpha$ number of floes, each a factor $\alpha^{-1}$ slower than the incident floe, so the energy is lowered by a factor $\alpha \times \alpha^{-2} = \alpha^{-1}$. Equating the energies we estimate that $\alpha = 2/(1+r^2)$. This prediction is consistent with both the perfectly elastic ($r = 1$) and perfectly inelastic ($r = 0$) limits. At $r = 1$, we obtain $\alpha = 1$, leading to $Q = 0$ as expected. Meanwhile, at $r = 0$, we predict $\alpha = 2$: a typical collision generates two energetic floes, each with half velocity of the incident floe.

*Limit of large collision rates*

We analyze the BTE for large collision rates, focusing on the tails of the distribution. Since $P(v_\perp)$ decays exponentially or faster and $\alpha > 1$, we infer that $\alpha P(\alpha v_\perp) \ll P(v_\perp)$, and drop it from the BTE. We solve the resulting equation using a WKB procedure [39], seeking a solution of the form $P(v_\perp) = Z \exp\{f(v_\perp)\}$, where $Z$ is a normalization factor that ensures that $\int_{-\infty}^{\infty} P(v_\perp) dv_\perp = 1$. Substituting this ansatz into the simplified BTE gives

$$\mathcal{N}^2 \sigma^2 (f''(v_\perp) - f'(v_\perp)^2) + v_\perp f'(v_\perp) = 1 - \nu\tau, \quad (15)$$

where the prime denotes a derivative with respect to the argument. To identify a dominant balance of terms we first rescale $v_\perp = \mathcal{N}\sigma V$ and $P(v_\perp) = \gamma F(V)$, where $V$, $F$ are dimensionless variables, and $\gamma$ is a dimensionless scaling factor that needs to be picked. Substituting yields

$$\gamma F''(V) - \gamma^2 F'(V)^2 + \gamma V F'(V) = -(\nu\tau - 1). \quad (16)$$

The dominant term on the left is the one scaling as $\gamma^2$, which comes from wind fluctuations, and must balance the right-hand side, which is due to collisions. This dominant balance of terms yields $\gamma = \sqrt{\nu\tau - 1}$ and $F'(V)^2 \approx 1$, leading to the solution $F(V) = -|V|$. Reverting to dimensional variables, we obtain

$$P(v_\perp) \sim \frac{q}{2} e^{-q|v_\perp|}, \quad \text{with} \quad q = \frac{\sqrt{\nu\tau - 1}}{\mathcal{N}\sigma}. \quad (17)$$

### D. Theoretical estimates of the collision rate, RMS velocity and diffusivity

We assume a power law distribution of floe sizes with exponent $-2$, truncated at inner and outer limits of $a_1$ and $a_2$ (10 m and 4500 m, respectively). We define the size distribution function

$$P_s(a) = \frac{a_2 a_1}{a_2 - a_1} a^{-2} \quad (a_1 \leq a \leq a_2), \quad (18)$$

such that $P_s(a) da$ is the probability that a randomly selected floe has a radius between $a$ and $a + da$; note that $\int_{a_1}^{a_2} P_s(a) da = 1$. The momentum of a floe depends on its mass (which scales with area), while the rate of exchange of momentum due to collisions depends on the number of neighbors, which scales with the perimeter. The average floe radius that preserves the perimeter-to-area ratio of the full distribution is

$$\bar{a} = \frac{\int_{a_1}^{a_2} a^2 P_s(a) \, da}{\int_{a_1}^{a_2} a P_s(a) \, da} = \frac{a_2 - a_1}{\log(a_2/a_1)}. \quad (19)$$

To estimate the mean-free-path we imagine a square lattice of floes with radius $\bar{a}$ separated by distance $\lambda$ (we use a square lattice for simplicity but other geometries yield similar estimates). The ice area fraction is then $\varphi = \pi \bar{a}^2/\lambda^2$, leading to

$$\lambda = \sqrt{\frac{\pi}{\varphi}} \bar{a} \quad (20)$$

Interpreting $\lambda$ as the mean-free-path and assuming an RMS ice velocity of $\mathcal{N}\sigma$ yields a typical collision time

$$\tau_{\text{coll.}} = \frac{\lambda}{\mathcal{N}\sigma}. \quad (21)$$

The collision rate is $\nu \approx \tau_{\text{coll.}}^{-1} = \mathcal{N}\sigma/\lambda$.

Using these theoretical estimates in the kinetic theory (5), we find an RMS velocity

$$\langle v_\perp^2 \rangle^{1/2} = \sqrt{\frac{2}{\nu\tau - 1}}\mathcal{N}\sigma \approx \sqrt{\frac{2}{\mathcal{N}\sigma\tau/\lambda - 1}}\mathcal{N}\sigma, \quad (22)$$

and, from (6), a diffusivity

$$D = \frac{\langle v_\perp^2 \rangle}{\nu} = \frac{2\mathcal{N}^2\sigma^2}{\nu(\nu\tau - 1)} = \frac{2\mathcal{N}\sigma\lambda}{\mathcal{N}\sigma\tau/\lambda - 1}, \quad (23)$$

both valid for large collision rates ($\nu\tau \gg 1$).

For the parameters used in the SDEM model and for $\varphi = 0.9$, we obtain an effective radius $\bar{a} \approx 735$ m, a mean-free-path $\lambda = 1375$ m, and $\tau_{\text{coll.}} = 3.2$ hr. It is noteworthy that monodisperse simulations with a radius of 750 m – close to the estimated $\bar{a}$ – closely approximate the results of the polydisperse simulations [6]. Additionally, the estimated collision time $\tau_{\text{coll.}} \approx 3.2$ hr is close to the best-fit time $\nu^{-1} = 3.7$ hr of the kinetic theory. Using the estimated $\nu$ (without any fitting parameters) yields $\langle v_\perp^2 \rangle^{1/2} \approx 0.055$ m/s and a diffusivity $D \approx 35 \text{ m}^2/\text{s}$, both quite close to the simulated results. Using the best-fit $\nu$ instead brings the theoretical prediction to within about 5% of the simulations, as discussed in the main text.

# Wind-driven collisions between floes explain the observed dispersion of Arctic sea ice
## Supplementary Information

Bryan Shaddy,[1] P. Alex Greaney,[2] and Bhargav Rallabandi[2, *]

[1]*Department of Aerospace and Mechanical Engineering,*
*University of Southern California, Los Angeles, California 90089, USA*
[2]*Department of Mechanical Engineering, University of California, Riverside, California 92521, USA*

## I. ERA5 WIND DATA ANALYSIS

To obtain wind statistics for the Fram Strait region to inform input values for our model, the 10 m wind components $(u_x, u_y)$ for the Fram Strait region, running in the E-W and N-S directions respectively, were analyzed for the winter months of January, February, and March for the years 2002, 2003, 2007, 2008, and 2009, matching the time periods considered for the analysis of FRAMZY/ACSYS ice buoys presented in [1]. Reanalysis data for the wind components over the Fram Strait (latitudes: 77° to 81°; longitudes: −5° to +5°) were acquired from the ERA5 dataset [2]. Averaging the wind components over the spatial coordinates and times considered, average wind speeds were computed for the $u_x$ and $u_y$ directions, from which an average wind direction was computed (stream-wise direction), along with the corresponding perpendicular direction (cross-stream). The $(u_x, u_y)$ wind components were projected onto the new stream-wise $u_\parallel$ and cross-stream $u_\perp$ coordinate directions, following which the mean wind velocities in these directions are subtracted to compute the wind velocity fluctuations in the two new coordinates. The standard deviation of wind velocity fluctuations in the two coordinate directions is then determined, along with the correlation time for the wind speed fluctuation time-series based on its autocorrelation function (ACF). The ACF of a stochastic time series $u(t)$ is defined by

$$\mathrm{ACF} = \int u(s)u(s+t)ds. \tag{S1}$$

The wind speed fluctuation standard deviation and correlation times for the stream-wise and cross-stream directions were then averaged to determine the final standard deviation and correlation time to be prescribed for the isotropic wind fluctuations in the stochastic wind model (2) that drive the ice motion. Following this analysis, the values determined for the model include a mean wind speed of 4.8 m/s, a standard deviation of 6.0 m/s, and a correlation time of 32 hours. Figure S1 depicts the probability density function (PDF) for the wind velocity fluctuations for the $u_\parallel$ and $u_\perp$ directions, along with the corresponding autocorrelation functions.

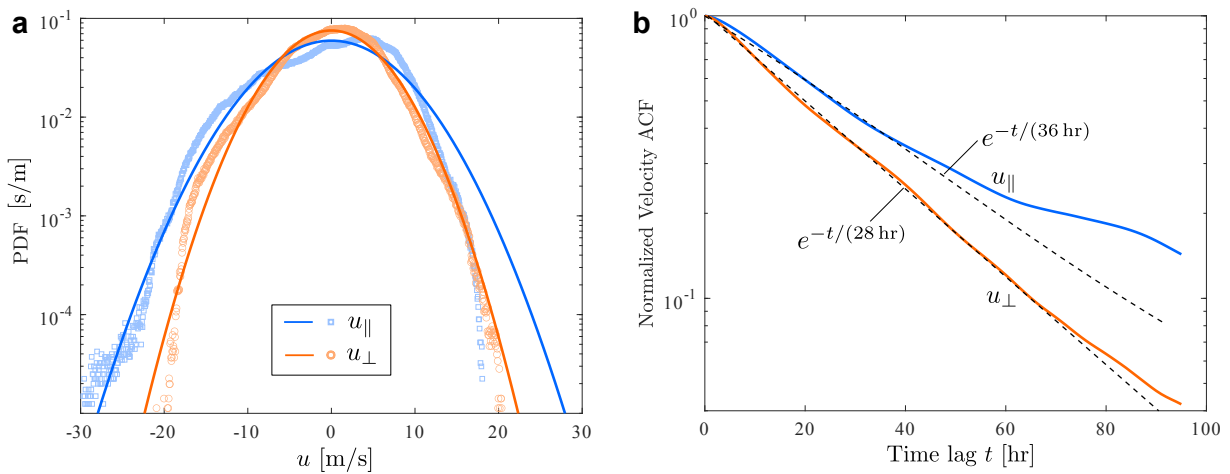

FIG. S1. (a) ERA5 $u_\parallel$ and $u_\perp$ wind velocity fluctuation PDFs with Gaussian fits overlaid and (b) ERA5 $u_\parallel$ and $u_\perp$ wind velocity fluctuation ACFs.

---

* bhargav@engr.ucr.edu

## II. ADDITIONAL MODEL INPUT PARAMETERS

Parameters characterizing the physical sea ice system for the simulation results presented here are listed in Table I [3, 4]. The ice thickness was obtained by computing the mean of the thickness distribution analyzed and modeled in [5].

| Parameter | Symbol | Value |
|---|---|---|
| Ice density | $\rho$ | 910 kg/m$^3$ |
| Ice thickness | $h$ | 2.2 m |
| Air density | $\rho_a$ | 1.3 kg/m$^3$ |
| Air drag coefficient | $C_a$ | $1.5 \times 10^{-3}$ |
| Water density | $\rho_w$ | 1025 kg/m$^3$ |
| Water drag coefficient | $C_w$ | $5 \times 10^{-3}$ |

TABLE I. Input parameters to model.

## III. PARAMETER STUDY

To examine the dependence of dispersion statistics on input parameters for which a range of values are observed or where a single "best estimate" value has not been empirically determined, simulations were run for a range of values. The parameters considered are the ice area fraction $\varphi$, which can vary over the course of a sea ice trajectory, and the coefficient of restitution $r$, for which an exact value is not known. We run simulations using the same parameter values as used for simulations presented in the main text, the only difference being the varied parameter. The area fraction covered by ice, $\varphi$, is studied in the main text. We study a range of values between $0.84 - 0.92$, accounting for much of the range observed in [1] while avoiding a jammed state. We further examine smaller values of $\varphi$ (down to 0.01), to gain a more complete picture of its effect on ice dynamics. The coefficient of restitution $r$ can range in $[0, 1]$ with 1 representing an elastic collision; here we consider $r$ in $[0.1, 0.9]$. From these simulations the diffusivity was computed and examined against $\varphi$ and $r$, for which plots are presented in Figure S2. We compute diffusivity in two different ways, (i) from a fit to the long-time dispersion $\langle x_\perp^2 \rangle \sim 2Dt$, and (ii) from the power spectral density at zero frequency: $D = \frac{1}{2}\text{PSD}(f=0)$. Both methods yield very similar results.

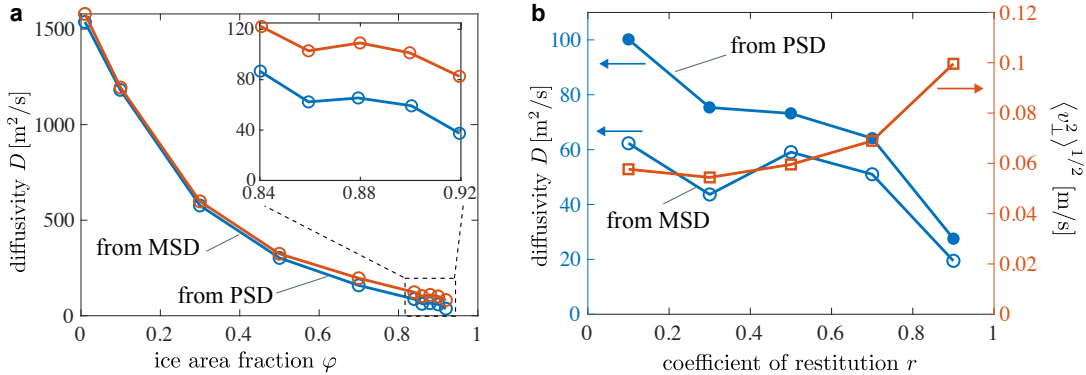

FIG. S2. Dispersion coefficient values (computed from both mean-squared-displacement (MSD) and power spectral density (PSD)) for polydisperse ice floes with varying input parameters. (a) $D$ versus $\varphi$ for a fixed $r = 0.1$. The inset shows the same data in the range $\varphi \in [0.84, 0.92]$, characteristic of the observational data analyzed in [1]. (b) $D$ versus $r$ for a fixed $\varphi = 0.9$.

For the range of $\varphi$ considered, we found that the dispersion coefficient diminished as the area fraction increased, with a value of 0.92 appearing to be just shy of the jammed state. When varying $r$ at fixed $\varphi = 0.9$, we find that the dispersion coefficient does not vary significantly for $r$ values below 0.7. Above this value, we find that the dispersion coefficient diminishes slightly, possibly due to the shorter mean-free times. The RMS velocity is also relatively insensitive to $r$ for small values of $r$ but increases somewhat as $r$ approaches 1. This can be understood by noting that less energy is dissipated by collisions for larger $r$, so the floes fluctuate more rapidly. It is interesting to note that the RMS velocity appears to approach about 0.12 m/s if the data is extrapolated to $r = 1$. This is consistent with the velocity of an isolated floe, as should be expected, since the energy of fluctuations is unaffected by collisions at $r = 1$. The same prediction is made by the kinetic theory, since the collision operator vanishes in this limit.

## IV. SIMULATIONS WITH A MONODISPERSE ICE FLOE SIZE DISTRIBUTION

To examine impact of the ice floe size distribution on dispersion statistics, simulations are repeated with uniformly-sized ice floes with a single "effective" radius of 750 m. We note that this radius is close to the area-to-perimeter preserving effective radius of the full distribution, calculated to be 735 m (see Methods). We also note that the maximum packing fraction of uniformly sized discs is $\approx 0.91$ (whereas for a power law distribution the packing fraction is close to 1). To account for this lower maximum packing density we choose a lower value of $\varphi = 0.64$ in the monodisperse simulations. All other input parameters are carried over from the polydisperse simulations discussed in the main text. Figure S3 depicts the resulting cross-stream MSD, along with the PDF and PSD for the cross-stream velocity fluctuations. The monodisperse results (with the adjusted $\varphi$) are clearly in good agreement with observational data [1]. We find that even in the case of a monodisperse floe size distribution, quantitatively similar dispersion statistics are obtained as those for the polydisperse case, exemplifying the relatively small importance of the floe size distribution. Similarly, the other hallmark qualitative features of the ice statistics are also well captured, including the non-Gaussian velocity distributions and the shape of the power spectrum.

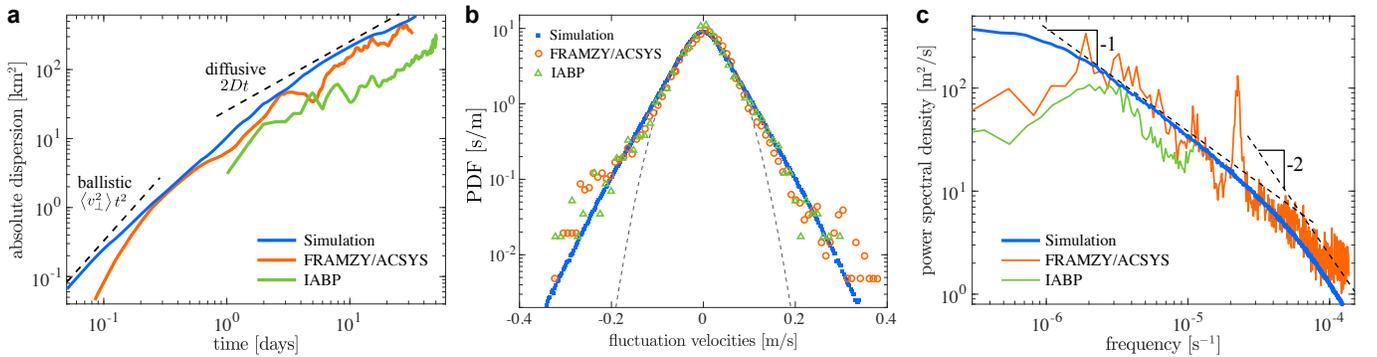

FIG. S3. Dispersion statistics for uniformly-sized ice floes with a radius of 750 m and area fraction of 0.64. (a) Cross-stream MSD. (b) PDF for cross-stream velocity fluctuations (dashed curve indicates a Gaussian fit). (c) PSD for cross-stream velocity fluctuations.